# Complete photonic band gaps in 3D foams

Ilham Maimouni*[a], Maryam Morvaridi*[a], Maria Russo*[a,d], Gianluc Lui[b], Konstantin Morozov[c], Janine Cossy[d], Marian Florescu[b], Matthieu Labousse[e], Patrick Tabeling[a]

[a]Microfluidique, MEMS et Nanostructures, Institut Pierre-Gilles de Gennes, CNRS UMR 8231, ESPCI Paris and Paris Sciences et Lettres (PSL) Research University, 75005 Paris, France.

[b]Advanced Technology Institute and Department of Physics, University of Surrey, Guildford, Surrey GU2 7XH, United Kingdom

[c]Department of Chemical Engineering Technion – Israel Institute of Technology Haifa 3200003, Israel

[d]Molecular, Macromolecular Chemistry and Materials, ESPCI Paris, CNRS, PSL University, 10 Rue Vauquelin, 75231 Paris Cedex 5, France

[e]Gulliver, CNRS UMR 7083, ESPCI Paris and Paris Sciences et Lettres (PSL) Research University, 75005 Paris France.

*These authors equally contributed.

**Corresponding author:** matthieu.labousse@espci.psl.eu

ORCIDs:

Matthieu Labousse: 0000-0001-8733-5863

Maria Russo: 0000-0001-8981-629X

**Author Contributions**

IM, MR carried out the experimental work; MM, GL performed the simulations; JC contributed in the synthesis of materials; KM contributed in the research; IM, MM, MR, GL, MF, ML, PT interpreted the results; MF, ML, PT supervised the research results; IM, MR, MF, ML, PT prepared the manuscript.

**Abstract**

To-date, despite remarkable applications in optoelectronics and tremendous amount of theoretical, computational and experimental efforts, there is no technological pathway enabling the fabrication of 3D photonic band gaps in the visible range. The resolution of advanced 3D printing technology does not allow to fabricate such materials and the current silica-based nanofabrication approaches do not permit the structuring of the desired optical material. Materials based on colloidal self-assembly of polymer spheres open 3D complete band gaps in the infrared range, but, owing to their critical index, not in the visible range. More complex systems, based on oriented tetrahedrons, are still prospected. Here we show, numerically, that FCC foams (Kepler structure) open a 3D complete band gap with a critical index of 2.80, thus compatible with the use of rutile $TiO_2$. We produce monodisperse solid Kepler foams including thousands of pores, down to 10 µm, and present a technological pathway, based on standard technologies, enabling the downsizing of such foams down to 400 nm, a size enabling the opening of a complete band gap centered at 500 nm.

**Main**

Photonic Band Gap (PBG) materials[1-7] provide a promising platform for the realization of remarkable applications in the fields of optoelectronics, energy harvesting and storage or communication[8-15]. There is a considerable current interest in the fabrication of 3D complete PBG materials exhibiting band gaps in the visible range to enable these exciting applications, but existing technologies face major challenges in the realization of large-area, defect-free 3D periodic dielectric structures. The resolution of advanced 3D printing technology does not allow to fabrication of such materials and the current silica-based nanofabrication approaches do not allow the structuring of the right optical material[16-20]. Self-assembly provides a promising platform for the realisation of scalable and cost-effective 3D PBG materials[21]. Materials based on colloidal self-assembly of polymer spheres[22], yielding inverted FCC (face-centered cubic) structures, have a greater potential and are known to open 3D complete band gaps in the infrared range. Nonetheless, such structures cannot provide a full photonic band gap in the visible range because, according to PBG calculations, the critical refractive index needed in the most favourable, experimentally realizable configuration, is 2.92[16] while the maximum refractive index of the champion of the domain – rutile Titanium Dioxide ($TiO_2$)- is 2.9[23,24]. To-date, directly self-assembled 3D photonic crystals do not possess a full photonic band gap in the visible range. In this context, several groups have developed approaches based on assemblies of more complex structures[18-20,25], such as tetrahedrons with oriented bonds[20]. This approach has received much attention over the past few years but has still to face a number of major challenges, including scaling down and self-assembly control. Here, we consider a new class of materials – 3D solid foams -, that can be self-assembled without complexifying the "molecular" structure and whose potential in terms of PBG has been suggested recently in 2D[25]. We show here that 3D solid FCC foams (Kepler structure) open complete 3D band gaps, with a critical refractive index equal to 2.80. This value enables, for the first time, the fabrication of materials opening 3D complete band gap in the visible range with currently available materials. We further describe a microfluidic technique that allows to create it, thus, paving a route towards the fabrication of PBG foams in the visible range.

3D liquid foams are self-assembled structures enabling the coexistence of a liquid and a gas phase. Their mechanical stability results from the minimization of the surface tension at the Plateau border, a phenomenon stabilized by adding surfactants in the liquid phase[26]. A solid 3D foam can be formed by adding of a polymer solution and a crosslinker in the liquid phase and then drying up the system[27-29].

Monodisperse liquid foams are locally ordered, and decades of experiments have shown that two stable crystalline structures exist, depending on the liquid fraction[30]. At high liquid fraction the system self-organizes into a Face Centered Cubic Structure (FCC, Kepler structure) while at low liquid fraction, the foam is a Body Centered Cubic structure (BCC, Kelvin structure). Note that although these structures have been observed for decades, they do not achieve full minimization of the surface energy of the system. Twenty years ago, it was shown that Weaire-Phelan foams, whose basic unit is composed by eight pores, achieve a minimal surface energy 0.3% deeper than Kelvin structure[31-32]. This structure is reminiscent of Si clathrates whose band diagram exhibits complete band gaps[33-36]. However, from an experimental prospective, contrarily to Kelvin and Kepler structures, this foam does not form spontaneously and need a forcing to exist. This feature raises serious experimental challenges for their realization[37].

Kepler and Kelvin structures form a crystal that we model by a network of connected rods of diameter $d$ and dielectric constant $\epsilon$ embedded within a unit cell of size $a$. The remaining volume is occupied by air of dielectric constant $\epsilon_{\text{air}} = 1$. We numerically investigate the optical properties of the two structures represented in Fig. 1a (respectively Fig. 1c). The band structure is found by solving numerically Maxwell equations through eigenmode decompositions with periodic boundary conditions and with the spectral MPB method[38], two approaches which are found in excellent agreement (see Methods and Supplementary Information). We investigate the band structure as a function of the dimensionless frequency $f = \omega a/(2\pi c)$, with $\omega$ the frequency of the electromagnetic wave and $c$ the speed of light. The band diagram of the FCC structure in Fig 1b evidences a complete band gap with a gap-midgap ratio $\%\text{BG} = \frac{\text{band width}}{\text{mean } f \text{ in the BG}} = 8.95$. Similarly, we numerically compute the band structure of the BCC structure and observe, as shown in Fig. 1d, a complete BG opening $\%\text{BG}=3.35$. The crossing of bands in Fig. 1b and 1d indicates a triple degeneracy at the points of the Brillouin zone which is expected as a natural consequence of the three four-fold rotational symmetry of the unit cell[1]. Notably, for both structures, the band gap does not close along the standard paths of the Brillouin zone (upper edges of the band gap occur along edges of the irreducible Brillouin zone), a result that has been confirmed by Density of States (DOS) calculations (See Supplementary information).

We numerically explore in Fig 2a, the dependence of the gap-midgap ratio on the geometrical and optical parameters of the unit cell. Maxwell equations are scale invariant so that the band gap depends only on the dimensionless rod's slenderness $\tilde{d}$=d/a and on the optical index $n = \sqrt{\varepsilon}$ of the rods. For a given rod optical index, we vary the rod diameter and measure the corresponding %BG. For both structures we observe that above a critical index a large band gap opens for a pronounced range of rod slenderness. In both structures an optimal rod slenderness is found and is a function of the rod optical index. We also note that for a given index, the %BG is always larger for a FCC structure than for a BCC structure. Fig. 2a evidences an increase of %BG with the optical index and more importantly a critical optical index ($n^*$) below which the band gap closes. Thus, in Fig. 2b we investigate the evolution of the maximum %BG with the rods index for both the FCC and BCC structures. The %BG increases linearly with the optical index and a critical index threshold is evidenced $n^* = 2.80$ for FCC and $n^* = 3.21$ for BCC. It is noteworthy to remind that large optical index contrasts are usually required to open band gaps. For example, for the inverse opal structure[22], a band gap opens for optical index larger than 2.92. In contrast, Fig. 2b shows that FCC solid foam exhibits a critical index $n^* = 2.80$ which makes it possible to open a band gap using the rutile $TiO_2$ for example ($n = 2.9$), a key result of this investigation.

The mechanism of the band gap opening can be rationalized by observing the spatial distribution of the field mode intensity of the dielectric and the air band[1], the band respectively below and above the gap (See Supplementary Information Fig. S2). For the air band, the field intensity is concentrated into region of small dielectric constant[1]. Conversely, for the dielectric band the field intensity is concentrated into region of large dielectric constant. This field concentration into different dielectric areas is driven by the dielectric contrast between these two areas. Therefore, for large dielectric contrast the coexistence of the dielectric and air bands is precluded in the frequency domain and a band gap opens.

Finally, the question of the fabrication of the foam with the two different organizations (FCC and BCC) is addressed. To fabricate liquid foams, standard microfluidic methods can be leveraged[39-41]., Flow-focusing, where the liquid streams focus the gas jet through a tiny orifice, is one of the commonly used geometries. To get controlled solid foams, the use of a polymer aqueous solution as the liquid phase and a crosslinker allows the solidification of primarily liquid foam. Injecting the crosslinker in the polymer phase at the entry of the chip allows distributing it homogeneously in the final structure. However, the preserving of the liquid foam template through the solidification is not straight forward. In the case of solid foams, for microfluidic channels widths under 100 µm, this approach no longer works.

The reason is that, unlike when simple Newtonian liquids are used as continuous phases (simple liquid foams), channels get clogged due to unavoidable formation of crosslinked polymer aggregates. So far, the smallest pores achieved within a monodisperse solid foam were around 80 μm wide[42]. To obtain smaller pores, we control the temperature at each stage of the microfluidic Flow Focusing production, as sketched in Fig. 3a. The inhibition of the crosslinking inside the chip (by lowering the temperature) and the accelerated solidification of the foam once recovered permits the use of micrometric (<50 μm) microfluidic channels, allowing for the first time the fabrication of highly controlled monodisperse solid foams with pores as small as 10 μm (see Methods). In this process, the formation of the foam starts with a bubble-by-bubble production at low temperature (5-10° C) until the gas volume fraction overcomes the jamming point[43] above which the bubbles self-assemble in monodisperse and ordered patterns. At this stage, the liquid foam is recovered and placed at a high temperature (50-100 °C) above its crosslinking temperature (40 °C)[44] for 1-5 minutes to quickly freeze its structure by a sudden increase of viscosity. Depending on the targeted height of the foam, a layer-by-layer solidification procedure might be necessary at this stage (see Supplementary Information, Fig. S3). The obtained solid structure shares the topological and geometric features of the original liquid template. Typically, the Plateau Borders (PB) thicken in the close vicinity of the nodes but are of a constant diameter in most of the central region, i.e. on two-third of the PB length, within a precision of 15% (see Supplementary Information, Fig. S4). As a whole, the 3D solid foam (Fig. 3b) can thus be described as an ordered network of rods meeting at nodes as assumed in the simulations. We observe in Fig. 3c and 3d that for a filling fraction φ above 9%, a FCC-like structure is observed (Fig. 3c, φ=11%), for 5% < φ < 9% (Fig. 3d, φ=8%), a coexistence of BCC and FCC structures is obtained and for φ < 5%, BCC-like structures are obtained[28]. The BCC/FCC transition results from topological changes in the foam structure, leading to the loss of the BCC characteristic quadrilateral faces[32]. In order to fully explore the potential of foams for photonic applications, the electromagnetic response of both configurations was calculated by our simulations. Coincidentally and in striking contrast with the BCC phase, the FCC structure combines all the advantages: it can be obtained pure, it exhibits larger PBG and at lower critical optical index.

Using the described temperature-regulated process, the final pores size is varied by adjusting the channels dimensions, the flow pressures and the reactants concentrations. Fig. 4a-d show confocal images of real 3D solid foams with a pore size ranging from 400 μm down to 10 μm. All the designed foams pores size distributions follow a normal distribution with a typical polydispersity index (normalized size standard deviation) of few percent (2% in Fig. 4e) which makes them suitable for photonic applications. The structures in the confocal images of Fig. 3 and Fig. 4 are composed of ordered layers in the out-of-plane direction of depths typically larger than 7-8 layers, a number above the evanescent penetration depth (3.2 layers see Methods). Consequently, we infer that, with our material, as desired, the electromagnetic field vanishes out in the band gap within the first encountered layers. The obtained solid foams can be further downscaled by combining the process described here with the standard downscaling methods existing for liquid emulsions. This task is achievable with lithography techniques and can be facilitated by using step-emulsification configuration, which produces droplets down to 400 nm with high throughput[43]. 400 nm bubbles would lead to a band gap centered on 500 nm, as shown above. At this wavelength, the refractive index of rutile $TiO_2$ is close to 2.9. These numbers indicate that our foam technology has a serious potential to create a complete PBG material in the visible range.

In this paper, we numerically demonstrated that 3D foam exhibits photonic band gaps, whose critical refractive index (2.80 for the Kepler structure) enables its fabrication with existing materials. We proposed a microfluidic approach enabling the fabrication of these solid foams, solving issues that thus far have jeopardized their miniaturization. Guided by the experiments, we restricted ourselves to the study of observed structures, i.e. Kelvin and Kepler foams. Weaire-Phelan foams are more energetically favorable and are known to open significantly wide PBG[33-36] but, in practice, need a forcing element to be obtained, raising serious experimental challenges. The foams designed in the scope of this work already open up numerous applications in the terahertz and the far infrared domains. For obtaining Kepler foams opening PBG in the visible range, our devices will need to be further scaled down, as mentioned above, and the polymer forming the template will have to be substituted by rutile $TiO_2$, a process already discussed in the literature[20].

**Methods**

*Photonic Band Gap Simulation:* Band structure and photonic eigenmode are calculated using spectral methods with a planewave basis (MPB software package[36]). We employ a 128x128x128 spatial mesh and a corresponding planewave basis consisting 1283 k-vectors, which ensures full convergence of the band structure calculation results. The band structure calculation results shown above are fully confirmed through finite element package *Comsol Multiphysics* (ver.5.4) (see Figure S1).

*Estimation of the penetration depth of the FCC structure in the band gap:* At the bottom of the air band the penetration depth[1] $\delta \simeq \left(\frac{\alpha}{\Delta f}\right)^{-1/2}$ with $\alpha = \frac{\partial^2 f}{\partial k^2}$ the local curvature along a given wave vector path and $\Delta f \simeq 5 \times 10^{-2}$ corresponding to half of the bandgap width. The bottom of the air band occurs along the wave vector path Γ→ K and we estimate $\delta \simeq 3.2 \pm 0.1$ [unit cell unit].

<u>Preparation of the liquid phase:</u> An aqueous solution of chitosan, a biopolymer extracted from shrimp shells, at a concentration between 0.3 and 2 %w/v was prepared. Chitosan having a medium molecular weight of 0.2·10⁶ and 0.4·10⁶ g/mol, Sigma Aldrich was dissolved in milliQ water leading to an aqueous solution of chitosan. Acetic acid was also added at a concentration of 0.1 mol/L. Dissolution and homogenization were obtained by stirring for at least 2 days at room temperature. Lutensol AT25 (provided by BASF), a non-ionic surfactant, was added at a concentration of 4 g/L. The surfactant permits stabilization of the liquid/air interface. The crosslinking agent Glyoxal (40% in water, purchased from Sigma Aldrich) was added to the thus obtained liquid phase at a concentration 2-10 %v/v.

*Foam production:* Foam generation is carried out in Flow-Focusing microfluidic geometry. The device is made by standard soft photolithography and replica-molding techniques using polydimethylsiloxane (PDMS). The channels have a depth of 5-50 µm and a width of 5-300 µm depending on the desired pore size. The setup used for the generation of monodisperse liquid foam templates is shown in Fig. 3a. It contains two inlets, one for injecting the liquid phase, one for injecting the gas phase and one outlet to collect the foam. At the nozzle, where the gas phase meets the liquid phase, micro-bubbles of gas are encapsulated in the continuous liquid phase. The flows are pressure-driven at a pressure $p_{gas}$ and $p_{liq}$ (0.05 - 5 bar) using a MFCS™-EZ pressure controller system (Fluigent) connected to a nitrogen tap. The gas phase is nitrogen with traces of perfluorohexane in order to hinder Ostwald ripening. The temperature of the microfluidic chip is fixed at 4-6 °C in order to suppress or prevent cross-linking until the liquid foam is outside the microfluidic equipment. The thus obtained liquid foam layer is then heated to 70 °C during few minutes, during which crosslinking occurs. The in-situ foam generation is observed using an Optical Microscope (Leika) with a 4-20x scanning objective.

*Foam characterization:* To observe foams, the sample is placed on a microscope glass slide, and images are taken using Optical and Confocal microscope systems (Leica). Confocal transmission images are analyzed by using a Matlab code that allows studying the topology of the foam (pore size distribution, PB profile).

**Acknowledgement**

The authors thank L. Berthier, R. Carminati, C.M. Cejas, P. Chaikin, Z. Amara, W. Drenckhan, A. Leshansky, D. Pine, S. Torquato, E. Yablonovitch for enlightening discussions, MMN group for experimental insights, ESPCI, CNRS, IPGG. MICROFLUSA (funded by the European Union Horizon 2020 research and innovation programme under Grant Agreement No. 664823) for their support. MF and GL acknowledge support from EPSRC (United Kingdom) Strategic Equipment Grant No. EP/L02263X/1 (EP/M008576/1) and EPSRC (United Kingdom) Grant EP/M027791/1.

**Figures**

Figure 1

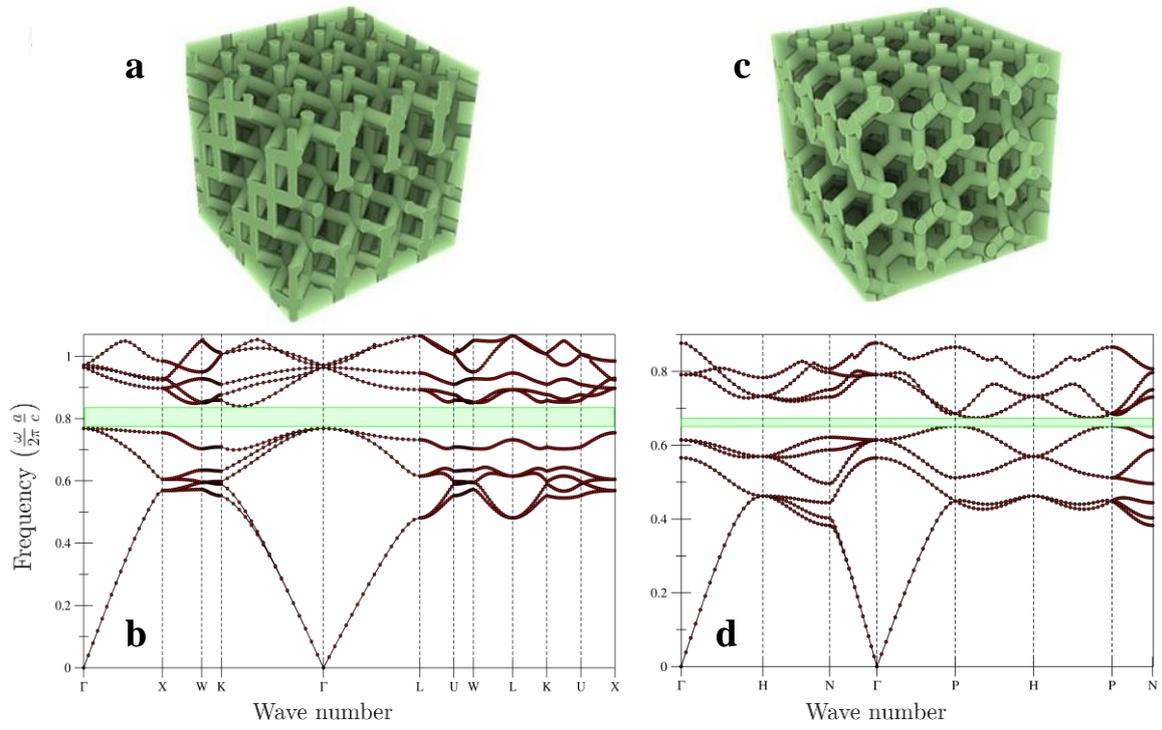

Figure 2

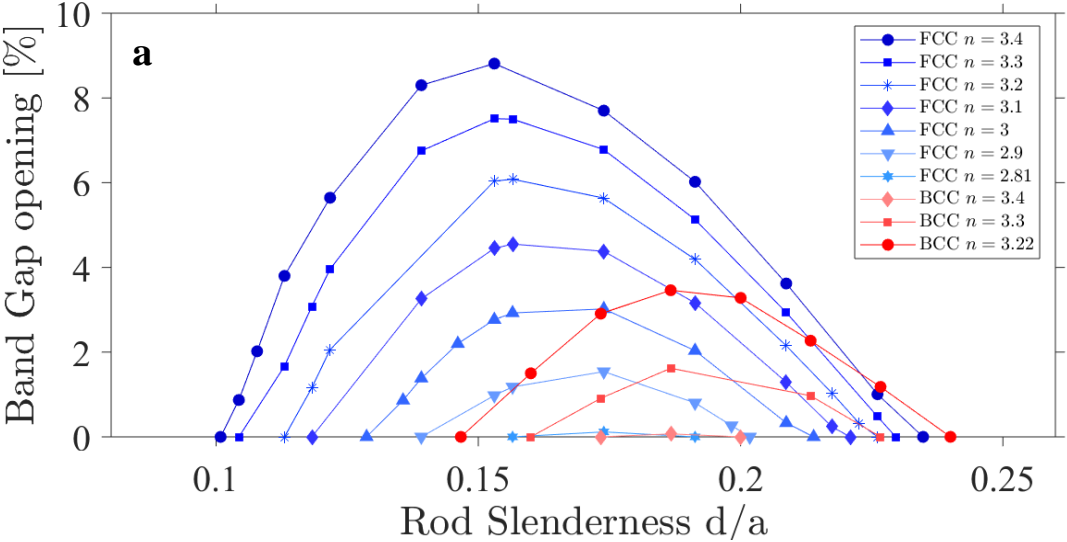

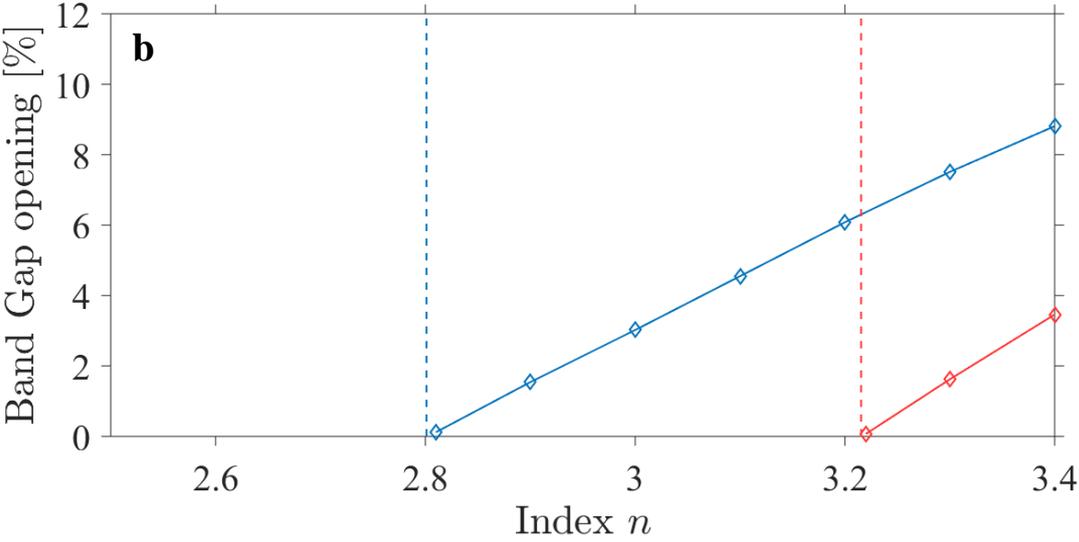

Figure 3

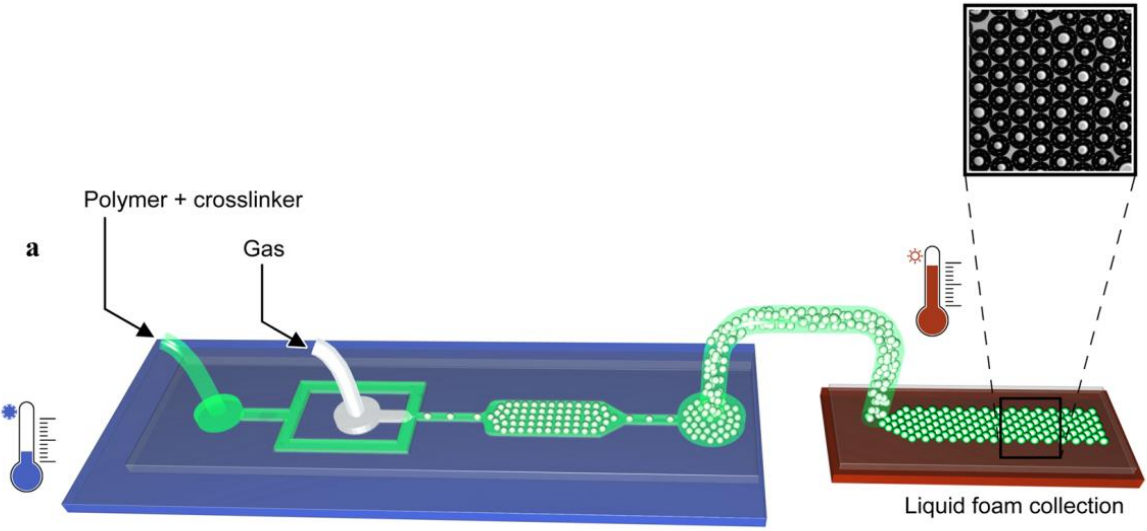

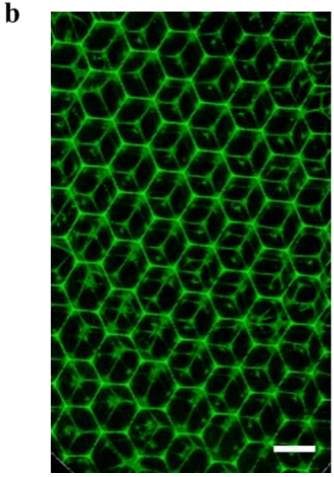
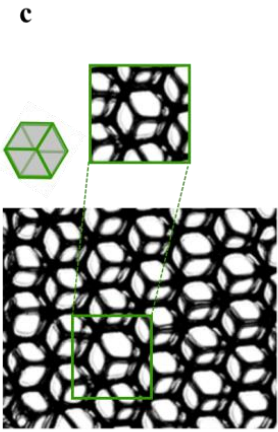
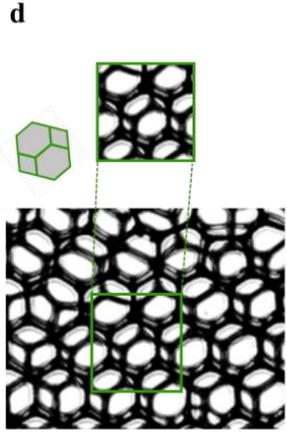

Figure 4

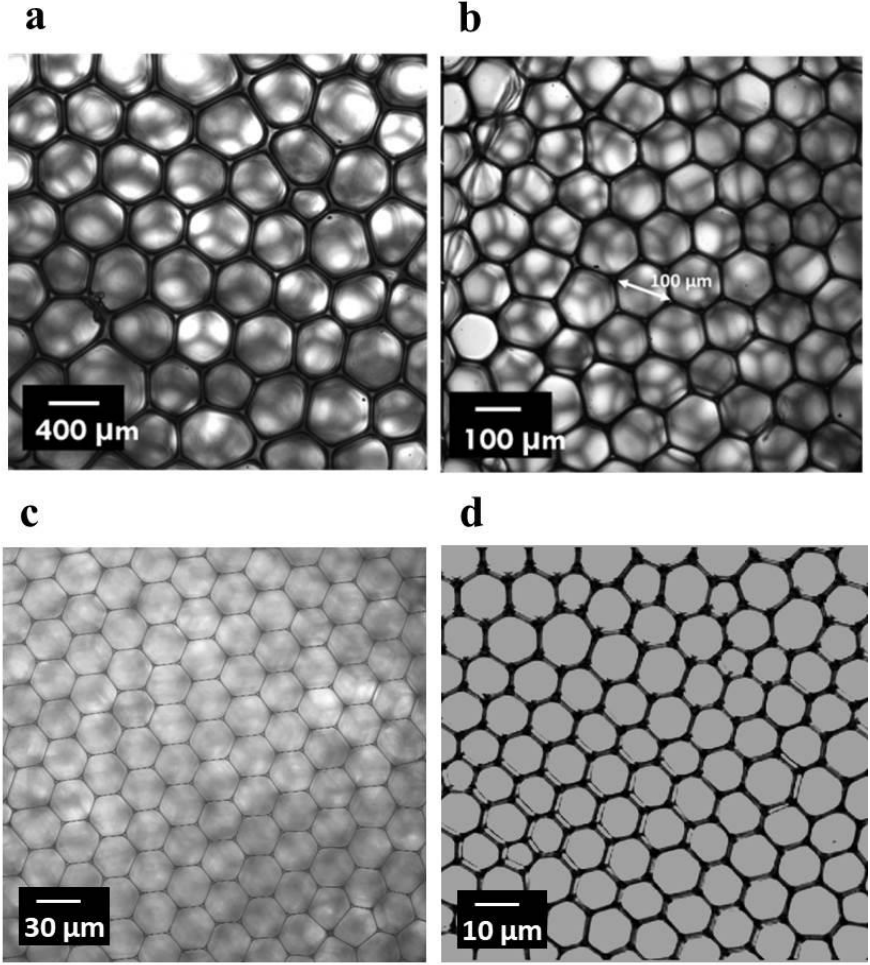

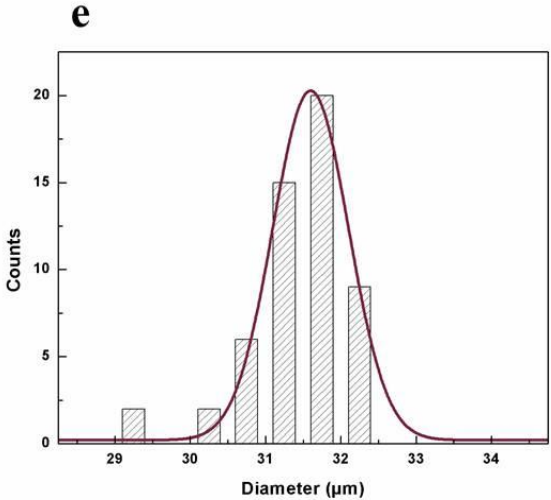

**Figure 1 Band structure calculations for the FCC and BCC structure. a** (respectively **c**) Schematic representation of the FCC structure (resp. BCC structure). **b** (resp. **d**) Band structure calculations FCC (resp. BCC) for the optimal structure comprising cylinders of diameter $d/a$ =0.153 (resp. $d/a$=0.202) corresponding to a filling ratio of 20.24% (resp. 20.94%), of relative dielectric constant $\epsilon = 11.56$ and a spatial mesh of 128x128x128. The structure displays a photonic band gap of 8.85% (resp. 3.28%) around a reduced central wave number $a/\lambda$=0.80 (resp. $a/\lambda$=0.66).

**Figure 2 Optimal geometry for photonic band gap opening. a** Numerical evolution of the gap-midgap ratio (%BG) with the dimensionless rods slenderness $\tilde{d} = d/a$. Red (resp. blue) colors refers to FCC (resp. BCC) structures. The darkening of the colors indicates larger optical indices n. **b** Numerical evolution of the geometrically optimal gap-midgap ratio (%BG) with the optical index of the rods. Vertical red (resp. blue) dashed line evidence the critical optical index $n^* = 2.80$ (resp. $n^* = 3.21$) to open a band in the FCC (resp. BCC) structure.

**Figure 3 Experimental realization of tunable targeted foam structures. a** Experimental sketch of the Flow-Focusing platform producing liquid foams – Insert of a liquid foam picture with characteristic roundish bubbles. **b** a typical 3D confocal reconstruction of the produced highly ordered open-cell solid foam with slender Plateau Borders of uniform thickness. Scale bar 100 µm. **c** FCC chitosan solid foam (Filling fraction 11%, pore diameter: 20 µm, Plateau border thickness ~ 11µm) **d** BCC solid foam crystal domain in a coexisting BBC-FCC phase., Filling fraction 8%, pore diameter: 20 µm, Plateau border thickness ~ 7µm).

**Figure 4 Experimental scaling down of the solid foam. a-d** Bright-field confocal images illustrating the pore size tuning between 400 µm and 10 µm with a narrow distribution. **e** The pores size distribution in **c** follows a Gaussian distribution, in the present case, centered around 31±0.6 µm (Polydispersity index 2%).

# Supplementary information: Complete photonic band gaps in 3D foams

Ilham Maimouni*[a], Maryam Morvaridi*[a], Maria Russo*[a,b], Gianluc Lui[c], Konstantin Morozov[d], Janine Cossy[b], Marian Florescu[c], Matthieu Labousse[e], Patrick Tabeling[a]

[a]Microfluidique, MEMS et Nanostructures, Institut Pierre-Gilles de Gennes, CNRS UMR 8231, ESPCI Paris and Paris Sciences et Lettres (PSL) Research University, 75005 Paris, France.

[b]Molecular, Macromolecular Chemistry and Materials, ESPCI Paris, CNRS, PSL University, 10 Rue Vauquelin, 75231 Paris Cedex 5, France

[c]Advanced Technology Institute and Department of Physics, University of Surrey, Guildford, Surrey GU2 7XH, United Kingdom

[d]Department of Chemical Engineering Technion – Israel Institute of Technology Haifa 3200003, Israel

[e]Gulliver, CNRS UMR 7083, ESPCI Paris and Paris Sciences et Lettres (PSL) Research University, 75005 Paris France.

*These authors equally contributed.

**Corresponding author:** matthieu.labousse@espci.psl.eu

**Band diagram using *Comsol Multiphysics*.**

MPB numerical results have been confirmed by using an eigenmode decomposition technique in *Comsol Multiphysics* (Ver 5.4). In this approach, the electromagnetic modes of photonic crystal are computed through eigenvalue problem with finite element package and applying Bloch's theorem. This approach has been validated for the case for the case of an FCC lattice of close packed dielectric spheres in air and the resulting band diagrams were in full agreement with the solution reported in the literature[1]. The important step to perform finite element method is assigning correct discretization. For 3D structures, the proper type of discretization is tetrahedral. The total volume of the system should be discretized in smaller elements. Finite element package *Comsol Multiphysics* is using proper optimization to get the solution convergence.

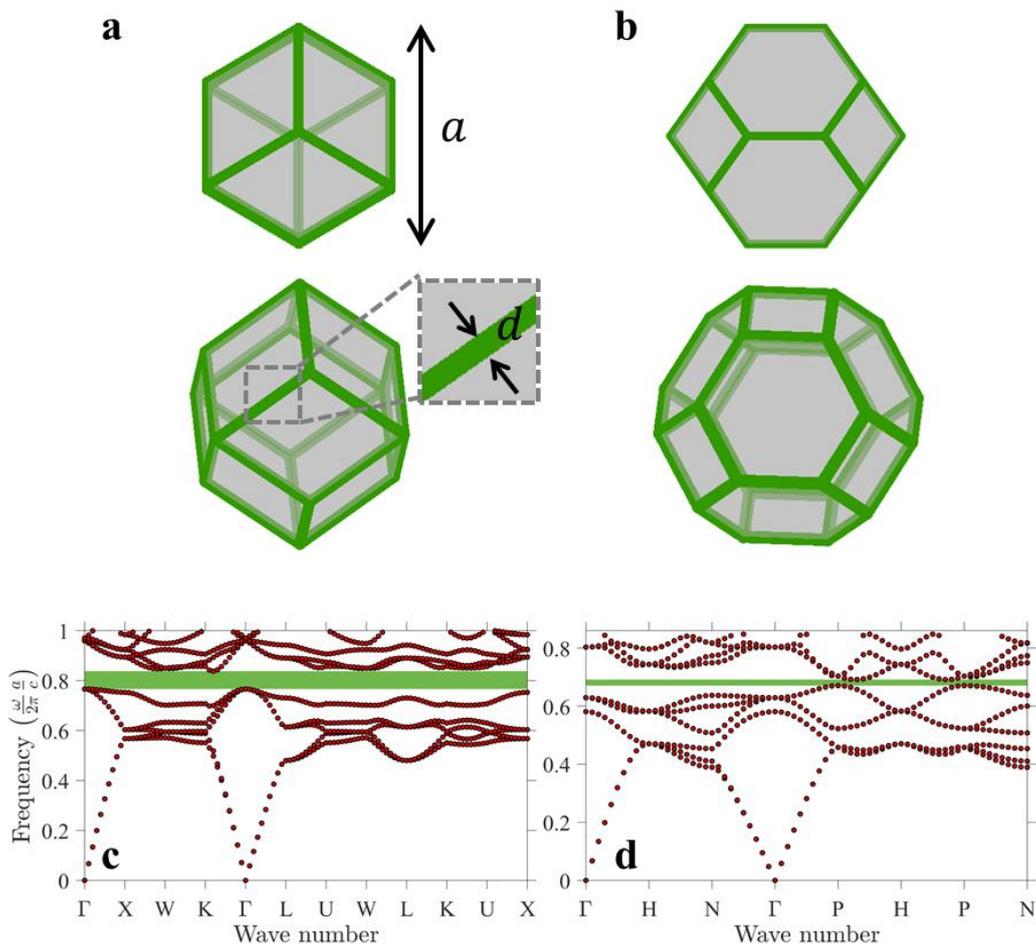

*Figure S.1. Solid foam structure and numerical photonic band diagram. a (respectively c) Geometry of the unit cell of the FCC-Kepler (resp. BCC-Kelvin) structure. Green rods of diameter d models a dielectric material while the remaining volume is air. b (respectively d) numerical band diagram of the FCC (resp. BCC) solid foam for $a = 115\,\mu m, d = 17.6\,\mu m, \tilde{d} = 0.153$ and $n = 3.4$ (resp. $a = 150\,\mu m, d = 30,27\,\mu m, \tilde{d} = 0.20$ and $n = 3.4$). Green bands indicate the band gap opening separating the lower dielectric band from the upper air band.*

**Numerical field distribution analysis.**

We illustrate in Figure S.2 the spatial field mode intensities for the optimal structures (n=3.4, $\tilde{d} = 0.153$ for the FCC structures and n=3.4, $\tilde{d} = 0.20$ for the BCC structures) at the scale of one unit cell (Comsol simulations, Figs. S.2a, S.2b, S.2e and S.2f), and for a finite sample consisting a few unit cells (MPB simulations, Figs. S.2c, S.2d, S.2g and S.2h). We display the spatial field mode intensity of the dielectric and air band at the Γ-point (FCC structure, Fig. S.2a and S.2b) and H-point (BCC, BCC structure, Fig. S.2e and S.2f). Similarly, we display the extended spatial profile intensity for the lower photonic gap band edges (Γ-point for the FCC structure in Fig. S.2d and the P-point for the BCC structure in Fig. S.2h) and for the upper photonic gap band edges (midway along the K-Γ direction for the FCC structure in Fig. S.2c and midway along the P-H edge for the BCC structure in Fig. S.2h). Accordingly to the electromagnetism variational theorem[1], the lower band gap edge has the field intensity concentrated into the region of larger dielectric constant (Figure S.2b and S.2d and Figure S.2f and S.2h) and conversely, the field intensity for the lower band edge is concentrated in the air fraction (Figure S.2a and S.2c and Figure S.2e and S.2g). This concentration of the field intensity into high/low index areas leads to the separation of the dielectric and air bands in the frequency domain and ultimately to the opening of the photonic band gap[1].

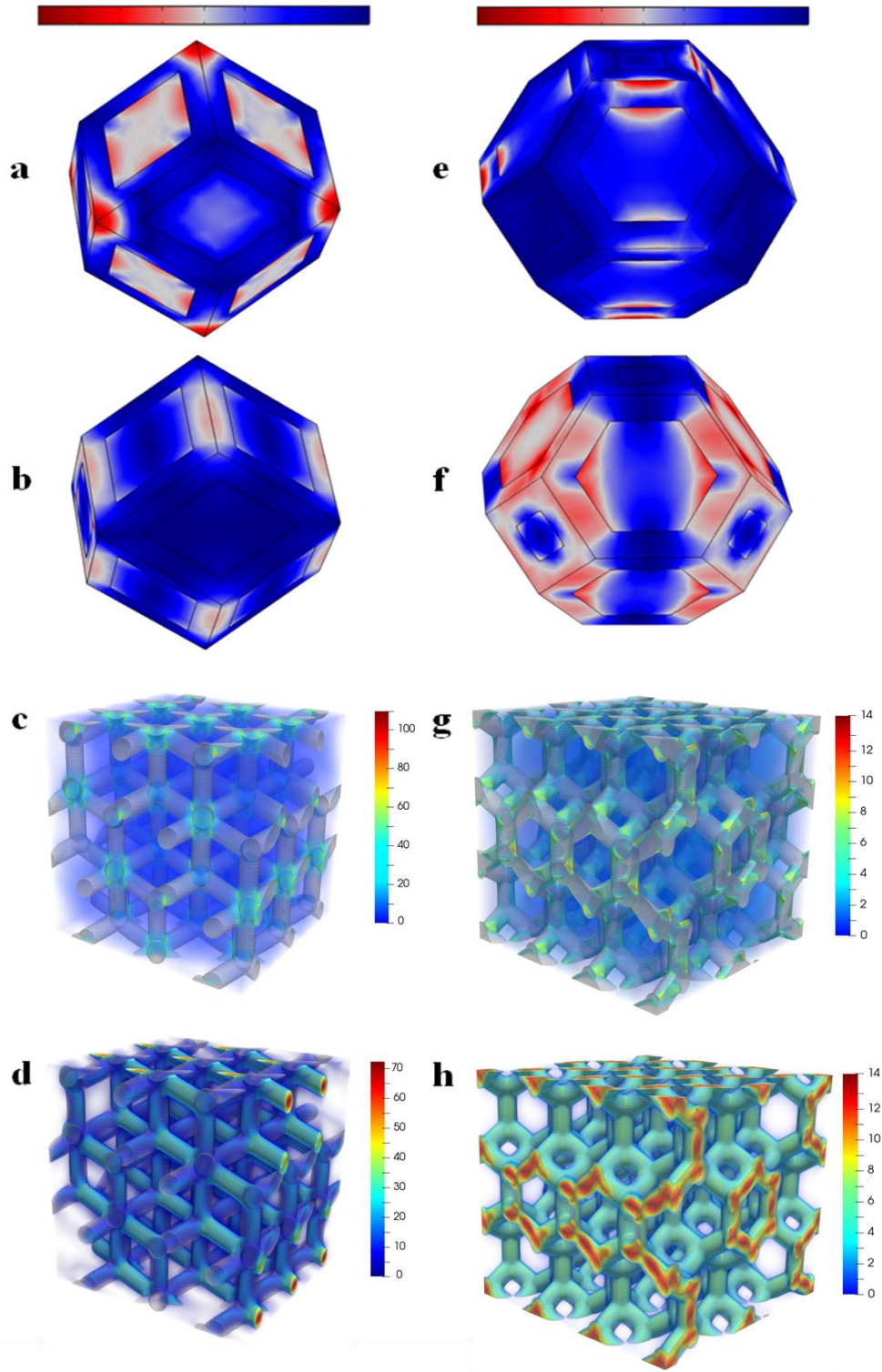

*Figure S.2. Analysis of field intensity when the gap band is open. a and b (resp e and f) Field intensity from Comsol calculations on a surface encapsulating the unit cell for FCC (resp. BCC) structure of geometry analogous to Fig. S1 at the Γ point of the Brillouin zone for FCC (H for BCC). The colormap accounts for the electric field norm intensity normalized by its maximum value (its significance being irrelevant for a linear system). a Air band of the FCC structure. b Dielectric band of the FCC structure. e Air band of the BCC structure. f Dielectric band of the BCC structure. c and d (resp. g and h) Field intensity from MPB calculations on a surface encapsulating the materials over several unit cells (resp. BCC). c Air band of the FCC structure (at its lower point). d dielectric band of the FCC structure (at its upper point). g Air band of the FCC structure (at its lower point). h Dielectric band of the FCC structure (at its upper point).*

**Layer-by-layer solidification process for foam sample higher than 1-3 mm.**

In order to minimize the evolution of structure during the solidification process due to the gravity-induced-drainage, the foam is generated following a "layer-by-layer" process: a first layer, of which the height is smaller than the capillary length of the liquid phase (around 1- 3 mm), is formed and sent to crosslinking. Before the crosslinking is complete (to ensure the adherence between the different layers), another layer is added above and the same operation is reiterated until the desired height of the sample is reached. Figure S.3 sums up the key steps that the foam goes through before we get the final samples.

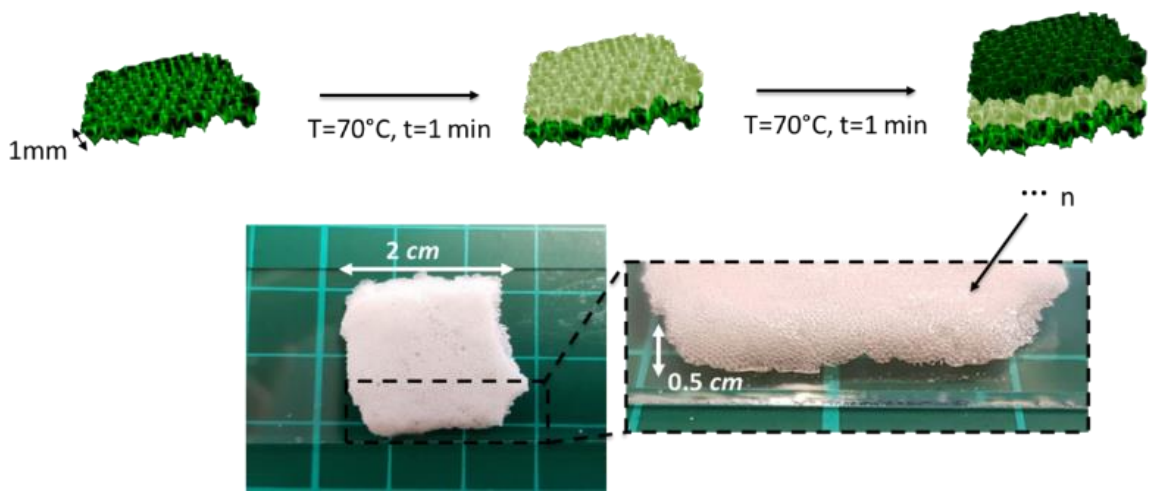

*Figure S.3. Layer-by-layer solidification process for foam sample higher than 1-3 mm: schematic representation of the different steps and a picture of an obtained chitosan foam (pore size 100µm)*

**Plateau Border thickness profile analysis.**

An image analysis of the Plateau Border (PB) diameter variation along its longitudinal axis (x axis) was performed using Matlab. First, the confocal image of a representative sample of a chitosan foam (pore size 50 µm) was thresholded (Figure S.4a). Then, the binary image was discretized into equally large strips along the *x axis* and the white area of each strip was calculated as a function of its coordinate. The graph represents the variation of the diameter of the PB along its x axis (Figure S.**4**.b). The results show that the plateau borders get thicker in the close vicinity of the nodes but are of a constant thickness in most of the central region (around two-third of the PB length).

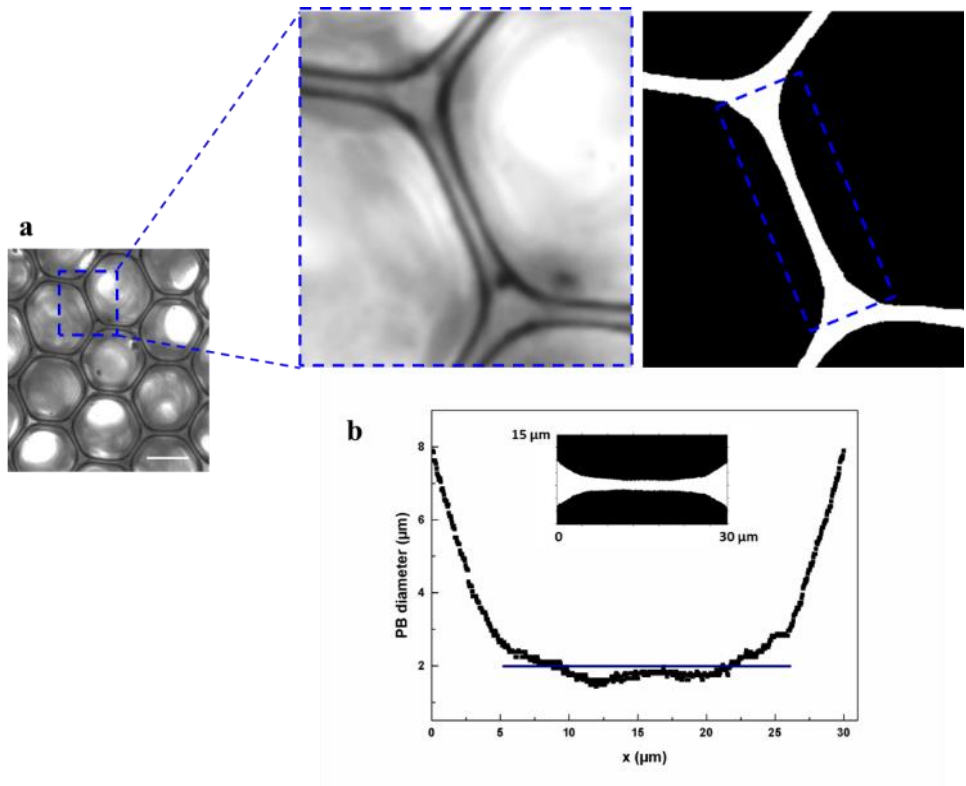

**Figure S.4** Plateau Border (PB) profile analysis. **a** Confocal bright-field image of chitosan solid foam with zoom on a Plateau Border and a contour profile. Scale is 30 µm; **b** Characteristic variation of PB diameter along its length.

**Density of States Calculations**

Both, the band structure for the FCC and BCC optimal structures (displayed in Fig.1 of the main text) have the upper edge of the photonic band gap falling midway along Brillouin zone edges[2], typically not included in conventional band structure calculations. Here, we confirm our predictions for the band gap size of the optimal FCC and BCC structures against predictions from density of states (DOS) calculations[3,4]. For DOS calculations, we use 32x32x32 k-points distributed uniformly in the Brillouin zone. As shown in Fig. S.5, we obtain a nearly perfect agreement between band structure and DOS predictions for the spectral location and size of the photonic band gap.

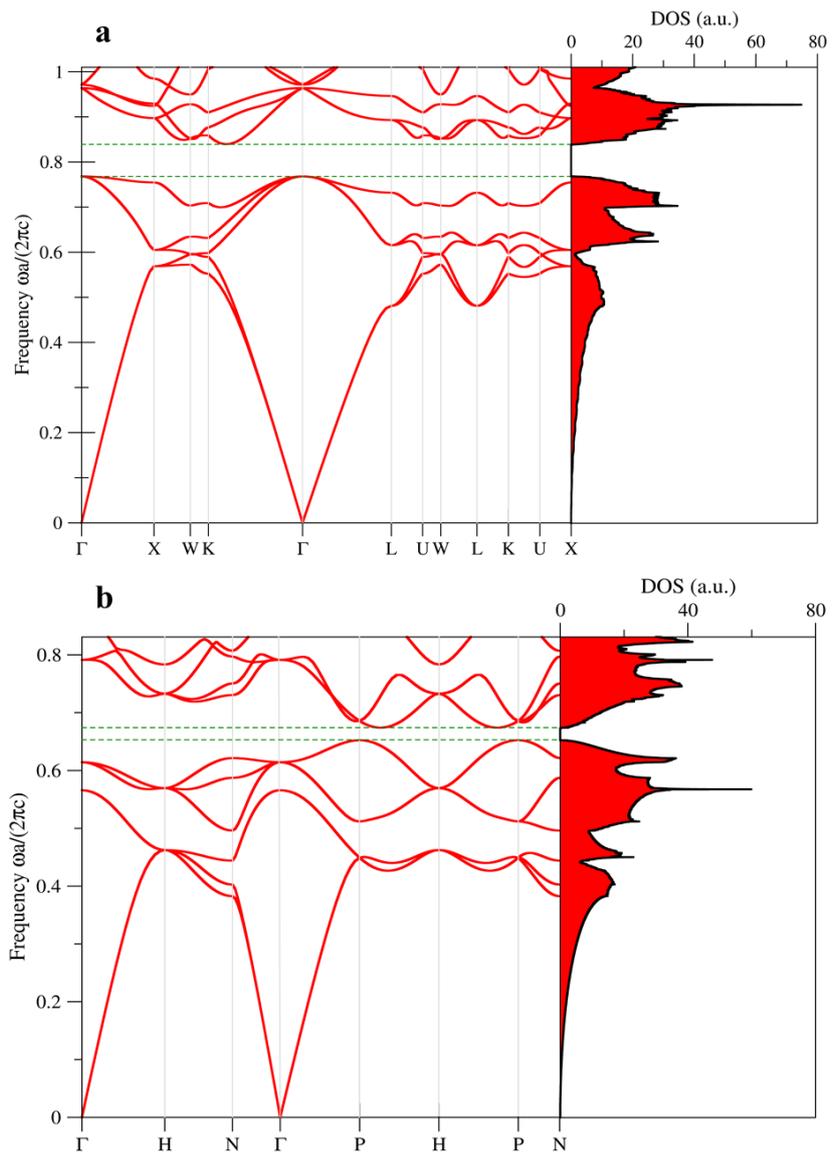

**Figure S5 DOS calculations. a** Band structure and DOS of the optimal FCC structure**. b** Band structure and DOS of the optimal BCC structure.